\documentclass[epj]{webofc}
\usepackage[varg]{txfonts}   
\usepackage{graphicx}
\usepackage{color,pst-plot}

\usepackage{amsmath}
\usepackage {amsfonts,amssymb,amstext}
\usepackage{psfrag}

\def\pv{\vec{p}_t}
\def\dv{\vec{\Delta}_t}

\def\ar{\alpha_\rho}

\def\mr{m_\rho}

\def\bea{\begin{eqnarray}}
\def\beqa{\begin{eqnarray}}
\def\eea{\end{eqnarray}}
\def\eqa{\end{eqnarray}}
\def\beas{\begin{eqnarray*}}
\def\eeas{\end{eqnarray*}}
\def\beqas{\begin{eqnarray*}}
\def\eqas{\end{eqnarray*}}
\def\beq{\begin{equation}}
\def\eeq{\end{equation}}
\def\beqd{\begin{displaymath}}
\def\eeqd{\end{displaymath}}
\def\eqd{\end{displaymath}}

\def\beeq{\begin{eqnarray}}
 \def\eeeq{\end{eqnarray}}
\newcommand{\eq}{\end{equation}}

\woctitle{Physics Opportunities at an Electron-Ion Collider}
\begin{document}
\title{Revealing transversity GPDs through the photoproduction of a photon and a $\rho$ meson}
%
%

\author{R. Boussarie\inst{1}\fnsep\thanks{\email{Renaud.Boussarie@th.u-psud.fr}}\and
B. Pire\inst{2}\fnsep\thanks{\email{bernard.pire@polytechnique.edu}} \and
        L. Szymanowski\inst{3}\fnsep\thanks{\email{Lech.Szymanowski@ncbj.gov.pl}} \and
        S. Wallon\inst{1,4}\fnsep\thanks{\email{Samuel.Wallon@th.u-psud.fr}}
}

\institute{Laboratoire de Physique Th\'eorique, UMR 8627, CNRS, Univ. Paris Sud, Universit\'e Paris-Saclay, \\ \mbox{}\hspace{.03cm} 91405 Orsay, France
\and Centre de physique th\'eorique,  \'Ecole Polytechnique, CNRS, Universit\'e Paris-Saclay, \\ 
\mbox{}\hspace{.03cm} 91128 Palaiseau, France
\and
           National Centre for Nuclear Research (NCBJ), Warsaw, Poland
           \and
           UPMC Univ. Paris 06, Facult\'e de Physique,75252 Paris, France }

\abstract{%
Photoproduction of a pair of particles with large invariant mass is a natural extension of collinear QCD
factorization theorems which have been much studied for deeply virtual Compton scattering and deeply virtual meson production. We discuss the production of a photon and a meson, where
the wide angle Compton scattering on a meson subprocess factorizes from generalized parton distribution.
We calculate at dominant twist and leading order in $\alpha_s$, the production cross-section of a transversely
polarized $\rho$ meson  which is sensitive to chiral-odd GPDs, and show that it may be measurable in near future JLab
experiments.
}
\maketitle
\section{Introduction}
Near forward photoproduction of a pair of particles with large invariant mass is a natural extension of collinear QCD factorization theorems which have been much studied for near forward deeply virtual Compton scattering (DVCS) and deeply virtual meson production~\cite{GPD}. We study here the reaction
\begin{equation}
\gamma^{(*)}(q) + N(p_1) \rightarrow \gamma(k) + \rho^0(p_\rho,\epsilon_\rho) + N'(p_2)\,,
\label{process1}
\end{equation}
 where a wide angle Compton scattering subprocess $\gamma (q\bar q) \to \gamma \rho $ characterized by the large scale $M^2_{\gamma \rho}=(p_\rho+k)^2$ (the final state $\gamma \rho $ invariant squared mass) factorizes from GPDs. This large scale $M_{\gamma \rho}$ is related to the large transverse momenta transmitted to  the final photon and to  the final meson, the pair having an overall small transverse momentum (since $\Delta^2 = (p_2-p_1)^2$ is small). 
The study of such reactions was initiated in Ref.~\cite{IPST}, where the process under study was the high energy diffractive photo- (or electro-)  production of two vector mesons, the hard probe being the virtual pomeron exchange (and the hard scale being the virtuality of this pomeron), in analogy with the virtual photon exchange occuring in the deep inelastic electroproduction of a meson.
The process we study here follows the same way of thinking as the one studied in Ref.~\cite{PLB}, there with a $\pi$ instead of a $\gamma$.

Finally, we note that a similar strategy has also been advocated in Ref.~\cite{kumano} to enlarge the number of processes which could be used to extract information on chiral-even GPDs. 

 The analogy to the timelike Compton scattering process~\cite{TCS} :
\begin{equation}
\gamma^{(*)} N  \to \gamma^* N' \to \mu^+ \mu^- N' \,,
\label{process2}
\end{equation}
where the lepton pair has a large squared  invariant mass $Q^2$, is instructive.  Although the photon-meson pair in the process  (\ref{process1}) has a more complex momentum flow, one may draw on this analogy to ascribe the role of the hard scale to the photon-meson pair invariant  mass.

The experimental study of such processes should not present major difficulties to modern detectors such as those developed for the 12 GeV upgrade of Jlab or for the Compass experiment at CERN. 
Probing various processes, thus verifying the universality of the GPDs, will  increase our confidence in a meaningful extraction of dominant (i.e. chiral-even) GPDs from near future experiments. 

 A noteworthy feature of the process  (\ref{process1}) is that, at dominant twist, separating the  transverse  (respectively longitudinal) polarization of the $\rho$ meson allows one to get access to  chiral-odd (respectively  chiral-even) GPDs. This is much welcome since access to the chiral-odd transversity GPDs~\cite{defDiehl}, noted  $H_T$, $E_T$, $\tilde{H}_T$, $\tilde{E}_T$, which decouple from deeply virtual Compton scattering and deeply virtual meson production at leading order, has  turned out to be most  challenging~\cite{DGP}.  Quark mass effects \cite{PS2015} or production of a meson described by a twist 3 distribution amplitude \cite{liuti} are two other ways which have been proposed to evade this difficulty. 

\begin{figure}[h]

\psfrag{TH}{$\Large T_H$}
\psfrag{Pi}{$\pi$}
\psfrag{P1}{$\,\phi$}
\psfrag{P2}{$\,\phi$}
\psfrag{Phi}{$\,\phi$}
\psfrag{Rho}{$\rho$}
\psfrag{tp}{$t'$}
\psfrag{s}{$s$}
\psfrag{x1}{$\!\!\!\!\!\!x+\xi$}
\psfrag{x2}{$\!x-\xi$}
\psfrag{RhoT}{$\rho_T$}
\psfrag{t}{$t$}
\psfrag{N}{$N$}
\psfrag{Np}{$N'$}
\psfrag{M}{$M^2_{\gamma \rho}$}
\psfrag{GPD}{$\!GPD$}
\centerline{
\raisebox{1.6cm}{\includegraphics[width=14pc]{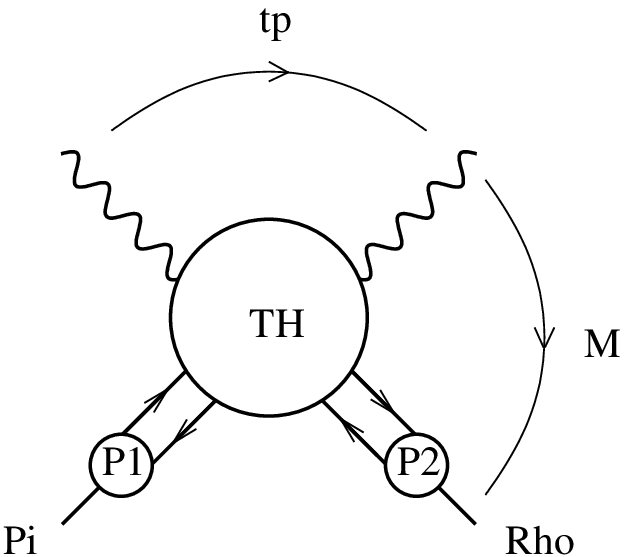}}~~~~~~~~~~~~~~
\psfrag{TH}{$\,\Large T_H$}
\includegraphics[width=14pc]{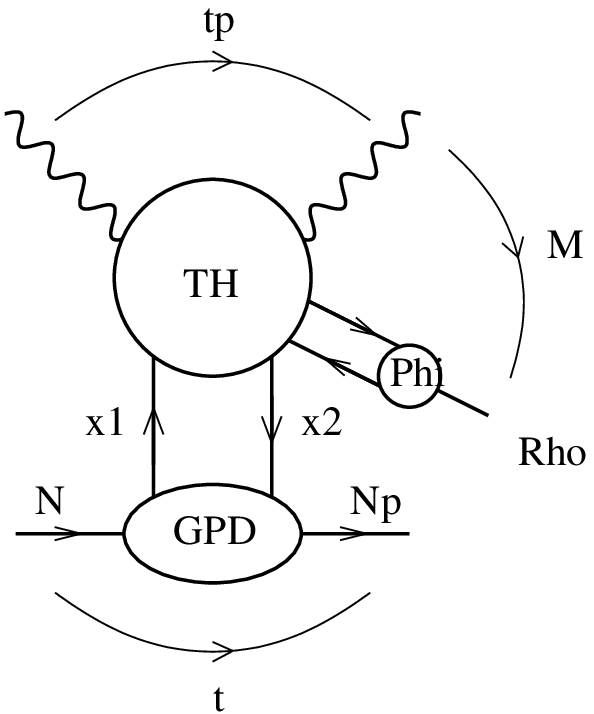}}
\caption{\label{Fig:process}The wide angle Compton scattering process (left) and its generalization to the photoproduction of a $\gamma \rho$ pair (right). }
\end{figure}

The calculation of the amplitude of process (\ref{process1}) relies on the use of the now classical proof of the factorization of exclusive scattering at fixed angle and large energy~\cite{LB}. The amplitude for the wide angle Compton scattering process $\gamma + \pi \rightarrow \gamma + \rho $ is written \cite{Nizic} as the convolution of mesonic distribution amplitudes (DAs)  and a hard scattering subprocess amplitude $\gamma +( q + \bar q) \rightarrow \gamma + (q + \bar q) $ with the final meson states replaced by a collinear quark-antiquark pair. 
In a second stage, we extract from the factorization procedure of the deeply virtual Compton scattering amplitude near the forward region the right to replace one entering meson DA by a $N \to N'$ GPD, and thus get Fig.~\ref{Fig:process} (right panel). 
The required skewness parameter $\xi$  is written in terms of the final photon - meson  squared invariant mass
$M^2_{\gamma\rho}$ as
\begin{equation}
\label{skewedness}
\xi = \frac{\tau}{2-\tau} ~~~~,~~~~\tau =
\frac{M^2_{\gamma\rho}-t}{S_{\gamma N}-M^2}\,.
\end{equation}

Indeed the same collinear factorization property bases the validity of the leading twist approximation which either replaces the meson wave function by its DA or the $N \to N'$ transition by nucleon GPDs. A slight difference is that light cone fractions ($z, 1- z$) leaving the DA are positive, while the corresponding fractions ($x+\xi,\xi-x$) may be positive or negative in the case of the GPD. Our  Born order calculation  shows that this difference does not ruin the factorization property.

In order for the leading twist factorization of a partonic amplitude to be valid, one should avoid the dangerous
kinematical regions where a small momentum transfer is exchanged in the
upper blob, namely small $t' =(k-q)^2$ or small $u'=(p_\rho-q)^2$, and the resonance regions for each  of the
invariant squared masses {$(p_\rho +p_{N'})^2,$} $(k+p_\rho)^2\,.$

Although we here limit ourselves to an initial on-shell photon, we note that our present discussion applies as well to the case of electroproduction where a moderate virtuality of the initial photon may help access the perturbative domain with a lower value of the hard scale $M_{\gamma \rho}$. 

\section{Kinematics}

We use the following conventions. We decompose all momenta on a Sudakov basis  as $
v^\mu = a \, n^\mu + b \, p^\mu + v_\bot^\mu $,
with $p$ and $n$ the light-cone vectors
$
p^\mu = \frac{\sqrt{s}}{2}(1,0,0,1), n^\mu = \frac{\sqrt{s}}{2}(1,0,0,-1), $
$v_\bot^\mu = (0,v^x,v^y,0) $ and $v_\bot^2 = -\vec{v}_t^2\,.
$
The particle momenta read
\begin{equation}
\label{impini}
 p_1^\mu = (1+\xi)\,p^\mu + \frac{M^2}{s(1+\xi)}\,n^\mu~, \quad p_2^\mu = (1-\xi)\,p^\mu + \frac{M^2+\vec{\Delta}^2_t}{s(1-\xi)}n^\mu + \Delta^\mu_\bot\,, \quad q^\mu = n^\mu ~,
\end{equation}
\beqa
\label{impfinc}
k^\mu = \alpha \, n^\mu + \frac{(\vec{p}_t-\vec\Delta_t/2)^2}{\alpha s}\,p^\mu + p_\bot^\mu -\frac{\Delta^\mu_\bot}{2},~ ~p_\rho^\mu = \alpha_\rho \, n^\mu + \frac{(\vec{p}_t+\vec\Delta_t/2)^2+m^2_\rho}{\alpha_\rho s}\,p^\mu - p_\bot^\mu-\frac{\Delta^\mu_\bot}{2},\nonumber
\eqa
with $\bar{\alpha} = 1 - \alpha$ and $M$,  $m_\rho$ are the masses of the nucleon  and the $\rho$ meson.
The total center-of-mass energy squared of the $\gamma$-N system is thus
\begin{equation}
\label{energysquared}
S_{\gamma N} = (q + p_1)^2 = (1+\xi)s + M^2\,.
\end{equation}
From these kinematical relations it follows that :
\beq
\label{2xi}
2 \, \xi = \frac{(\pv -\frac{1}2 \dv)^2 }{s \, \alpha} +\frac{(\pv +\frac{1}2 \dv)^2 + \mr^2}{s \, \ar}\,,
\eq
and
\beq
\label{exp_alpha}
1-\alpha-\ar = \frac{2 \, \xi \, M^2}{s \, (1-\xi^2)} + \frac{\dv^2}{s \, (1-\xi)}\,.
\eq
On the nucleon side, the transferred squared momentum is
\begin{equation}
\label{transfmom}
t = (p_2 - p_1)^2 = -\frac{1+\xi}{1-\xi}\vec{\Delta}_t^2 -\frac{4\xi^2M^2}{1-\xi^2}\,.
\end{equation}
The other various Mandelstam invariants read
\begin{eqnarray}
\label{M_pi_rho}
s'&=& ~(p_\gamma +p_\rho)^2 = ~M_{\gamma\rho}^2= 2 \xi \, s \left(1 - \frac{ 2 \, \xi \, M^2}{s (1-\xi^2)}  \right) - \dv^2 \frac{1+\xi}{1-\xi}\,, \\
\label{t'}
- t'&=& -(p_\gamma -q)^2 =~\frac{(\vec p_t-\vec\Delta_t/2)^2}{\alpha} \;,\\
\label{u'}
- u'&=&- (p_\rho-q)^2= ~\frac{(\vec p_t+\vec\Delta_t/2)^2+(1-\alpha_\rho)\, m_\rho^2}{\alpha_\rho}
 \; .
\end{eqnarray}
Let us remind that we are interested in the ``large angle'' kinematical domain where every invariant $s', -t', -u'$ is large (as compared to $\Lambda^2_{QCD}$), with the constraints $0 < \alpha, \alpha_\rho < 1$.

\section{The scattering amplitude}
\label{Sec:scattering}
In the spirit of the above described factorizations, the scattering amplitude of the process (\ref{process1}) is written as
\begin{equation}
\label{ampl}
\mathcal{A}(t,M^2_{\gamma\rho},u')  = \sum\limits_{q,i} \int_{-1}^1dx\int_0^1dz\ T_i^q(x,v,z) \, H_i^{q}(x,\xi,t)\Phi_{\rho_{L,T}}(z)\,,
\end{equation}
where
$T_i^q$ is the hard part of the amplitude and $H_i^{q}$ the corresponding (chiral-even and chiral-odd)
GPDs of a parton $q$  in the nucleon target, and $\Phi_{\rho_{L,T}}(z)$ the leading twist chiral-even (resp. chiral-odd) distribution amplitude of the $\rho_L$ (resp. $\rho_T$) meson.

\begin{figure}[h]
\centerline{\includegraphics[width=32pc]{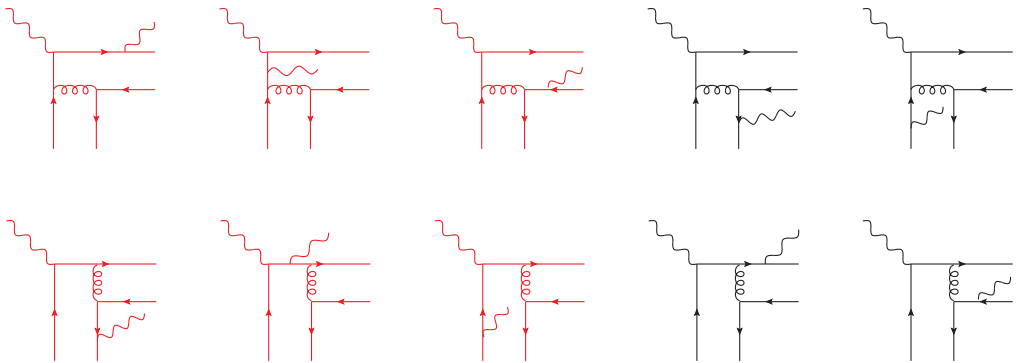}}
\caption{\label{Fig:diagrams}The Feynman diagrams describing the subprocess at leading order; in Feynman gauge only the 4 diagrams on the right contribute to the $\rho_T$ case}
\end{figure}
The scattering sub-process is described by
20 Feynman diagrams, but an interesting (quark-antiquark interchange) symmetry allows to deduce the contribution of half of the diagrams from the 10 diagrams shown on Fig.~\ref{Fig:diagrams} through a ($x \leftrightarrow -x \,; \,z \leftrightarrow 1-z$) interchange. Moreover, in Feynman gauge, only the 4 diagrams on the right of Fig.~\ref{Fig:diagrams} contribute to the chiral-odd case.

\begin{figure}[h]
\vspace{.5cm}
\psfrag{U}{\raisebox{-.15cm}{$1$}}
\psfrag{D}{\raisebox{-.15cm}{$2$}}
\psfrag{T}{\raisebox{-.15cm}{$3$}}
\psfrag{Q}{\raisebox{-.15cm}{$4$}}
\psfrag{C}{\raisebox{-.15cm}{$5$}}
\psfrag{a}{\raisebox{0cm}{\hspace{-.4cm}$0.5$}}
\psfrag{b}{\raisebox{0cm}{\hspace{-.4cm}$1$}}
\psfrag{c}{\raisebox{0cm}{\hspace{-.4cm}$1.5$}}
\psfrag{d}{\raisebox{0cm}{\hspace{-.4cm}$2$}}
\psfrag{e}{\raisebox{0cm}{\hspace{-.4cm}$2.5$}}
\psfrag{f}{\raisebox{0cm}{\hspace{-.4cm}$3$}}
\psfrag{g}{\raisebox{0cm}{\hspace{-.4cm}$3.5$}}
\psfrag{O}{\raisebox{-.15cm}{$0$}}
\psfrag{Z}{\raisebox{.3cm}{\hspace{-.9cm}$\left.\frac{d \sigma}{dt du' dM_{\gamma \rho}^2}\right|_{|t|_{\rm min}}$(pb.GeV$^{-6}$)}}
\psfrag{W}{\raisebox{-.3cm}{$-u'$ (GeV$^2$)}}
\centerline{\hspace{-.5cm}\includegraphics[width=12cm]{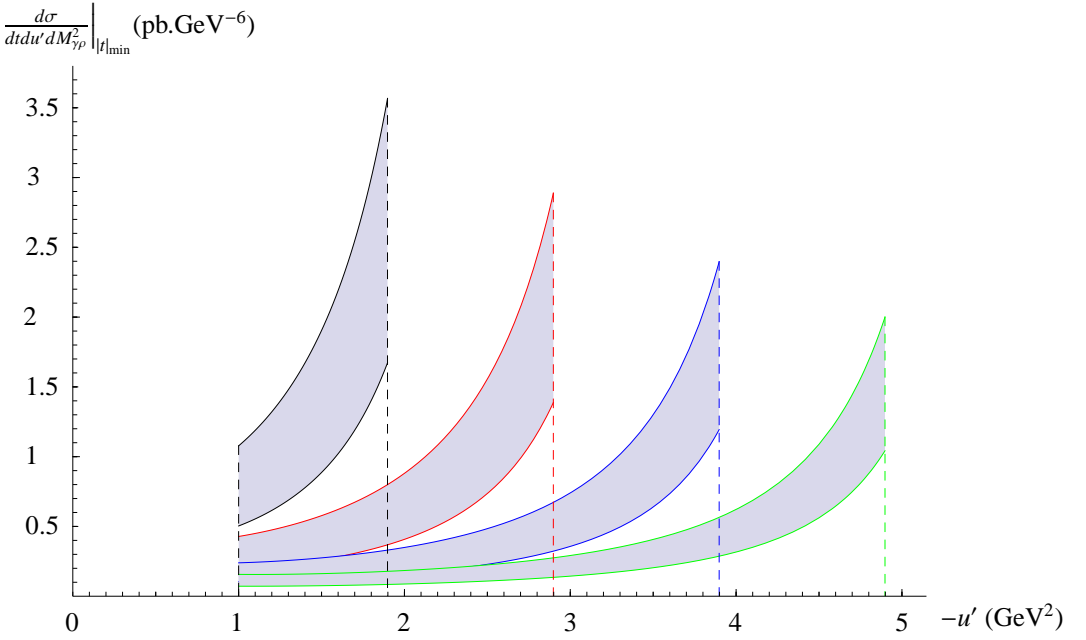}}
\vspace{.3cm}
\caption{The differential cross section (\ref{difcrosec}) for the production of $\gamma \rho_T$ involving chiral-odd GPDs, as a function of  $-u'$ at $S_{\gamma N}$ = 20 GeV$^2$  for (from left to right, resp. in black, red, blue, green) $M^2_{\gamma\rho}$ =  3, 4, 5, 6  GeV$^2$. The gray bands indicate the theoretical uncertainties.}
\label{Fig:resultS20}
\end{figure}

The scattering amplitudes get both  real and  imaginary parts. Focusing on the chiral-odd amplitude (since accessing transversity GPDs was the first motivation of our study), we get the following results. The $z$ and $x$ dependence  of this amplitude can be factorized as
\begin{equation}
 T_i^q = e_q^2 \,\alpha_{em}\, \alpha_s \,{\mathcal{N} (z,x)}\, \mathcal{T}^i
 \end{equation}
  with (in the gauge $p.\epsilon_k =0$):
\begin{eqnarray} 
\mathcal{T}^i &=& (1-\alpha) \left[ \left( \epsilon_{q\bot} . p_\bot \right) \left( \epsilon_{k\bot}.\epsilon_{\rho\bot} \right) - \left( \epsilon_{k\bot} . p_\bot \right) \left( \epsilon_{q\bot}.\epsilon_{\rho\bot} \right) \right] p_\bot^i \nonumber \\ 
&-& (1+\alpha) \left(\epsilon_{\rho\bot}.p_\bot\right) \left( \epsilon_{k\bot}.\epsilon_{q\bot}\right) p_\bot^i + \alpha \left( \alpha^2 -1\right) \xi s \left(\epsilon_{q\bot}.\epsilon_{k\bot}\right)\epsilon_\rho^i \\ 
&-&\alpha \left( \alpha^2 -1 \right) \xi s \left[ \left(\epsilon_{q\bot}.\epsilon_{\rho\bot}\right) \epsilon_{k\bot}^i - \left(\epsilon_{k\bot}.\epsilon_{\rho\bot}\right) \epsilon_{q\bot}^i \right]\,. \nonumber
\end{eqnarray}
Using as a first estimate the asymptotic form of the $\rho-$meson distribution amplitude, we perform analytically the integration over $z$. Inserting a model for the transversity GPDs \cite{PLB}, we use numerical methods for the integration over $x$.

Starting with the expression of the scattering amplitude (\ref{ampl}),
the differential cross-section as a function of $t$, $M^2_{\gamma\rho},$ $-u'$  reads
\begin{equation}
\label{difcrosec}
\left.\frac{d\sigma}{dt \,du' \, dM^2_{\gamma\rho}}\right|_{\ |t|=|t|_{min}} = \frac{|\mathcal{M}|^2}{32S_{\gamma N}^2M^2_{\gamma\rho}(2\pi)^3}\,.
\end{equation}
\noindent
We show in Fig.~\ref{Fig:resultS20}  this cross section (\ref{difcrosec}) as a function of  $-u'$ at $S_{\gamma N}$ = 20 GeV$^2$  for various fixed values of $M^2_{\gamma\rho}$ (namely 3, 4, 5, 6 GeV$^2$). There are rather huge uncertainties for these cross-sections, mainly of course because of our poor knowledge of  chiral-odd GPDs. One of the main uncertainties is due to the fact that the transversity PDFs~\cite{Anselmino:2013vqa}, on which we rely to build our model for transversity GPDs, are themselves not very constrained by existing data on semi-inclusive deep inelastic scattering.

The cross section grows with $(-u')$ but its normalization is rather small.  We expect a larger cross-section for the longitudinal $\rho$ case where chiral-even GPDs contribute; this will not help disentangling the transverse $\rho$ cross section. A complete analysis of the angular distribution of the emerging $\pi^+ \pi^-$ pair allows in principle to access the chiral-odd sensitive contribution at the amplitude level.
If such an analysis is performed, the high luminosity expected at JLab@12GeV might be sufficient for an experimental evidence of a needed contribution of transversity GPDs at twist 2, in the case of a $\rho_T$ in the final state.

\section*{Acknowledgments}

\noindent
L.~Sz was partially supported by grant of National Science Center, Poland, No. 2015/17/B/ST2/01838. This work is partly supported by the French grant ANR PARTONS (Grant No.ANR-12-MONU-0008-01).

\end{document}